\documentclass[10pt]{iopart}

\usepackage{graphicx}  
\usepackage{wrapfig}
\usepackage{epsfig}
\usepackage{subfigure}

\def\pt{$p_T$ }
\def\mt{$m_T$ }
\def\ee{$e^+e^-$ }

\def\gev2{GeV/$c^2$}

\def\mev2{MeV/$c^2$}

\begin{document}

\title[]{e$^+$e$^-$ pairs: a clock and a thermometer of heavy ion collisions}

\author{Alberica Toia}

\address{
Department of Physics and Astronomy,
Stony Brook University,
\\
Stony Brook, 11794-3800, NY USA
}
\ead{alberica@skipper.physics.sunysb.edu}

\begin{abstract}
Recently, there is growing evidence that a new state of
matter is formed in $\sqrt{s_{NN}}$ = 200 GeV Au+Au collisions at RHIC: a
strongly coupled Quark Gluon Plasma of partonic degrees of freedom
which develops a collective motion. Dilepton spectra are not
affected by strong interaction and can therefore probe the whole time
evolution of the collision. Thus they may be sensitive to onset
of deconfinement, chiral symmetry restoration, as well as the
production of thermal photons.
The PHENIX experiment measured the production of \ee pairs in p+p
and Au+Au collisions at $\sqrt{s_{NN}}$ = 200 GeV.
An enhanced dilepton yield in the
mass range 150$<m_{ee}<$750 MeV/c$^2$ is measured. The excess
increases faster with centrality than the number of participating
nucleons and is concentrated at \pt$<$1GeV/c. 
At higher \pt the excess below 300 MeV/c$^2$ has been related to an
enhanced production of direct photons possibly of thermal origin.
\end{abstract}

\section{The role of dileptons and photons in heavy ion collisions}
One of the main goals in nuclear physics is to investigate QCD and its properties
under extreme conditions of hot and dense matter. These
conditions are experimentally realized in relativistic heavy ion
collisions, which create a fireball
where the energy density by far exceeds typical hadronic values
and fundamental constituents of matter, quarks and gluons, are no
longer confined to color neutral hadrons.
Lattice QCD calculations indicate that this significant change in the
structure of matter coincides with the restoration of chiral
symmetry, which is linked to the origin of the hadron masses \cite{lattice}.
Experimental results from the Relativistic Heavy Ion Collider (RHIC)
have established the formation of such Quark Gluon Plasma in Au+Au collisions at
$\sqrt{s_{NN}}$=200GeV \cite{whitepaper}.
A large set of data allows to characterize
the nature of the medium created. 
It has very high density, as indicated by the large energy loss of light and
heavy quarks, and thermalizes rapidly, as indicated by the
significant elliptic flow of the same partons. Many open questions
remain, for example: what is
the temperatures, the corresponding system sizes of the matter at the
early stages, and  the temperature evolution of the
system from the formation time to thermal freeze-out? Is chiral
symmetry restored?
While signatures of chiral symmetry restoration manifest themselves
in non-observable order-parameters (such as the quark
condensate), links can be established with the hadronic excitations
and modifications of their fundamental properties, such as mass, width
and life-time.   

Electromagnetic probes, such as real and virtual photons
(i.e. dileptons), provide key measurements to address these open questions. Created through the entire space-time evolution of the
system, they escape, once emitted, from the strongly-interacting
medium without final-state interaction. 
The measurement of direct thermal radiation (photons or
dileptons) can be used to derive a limit on the initial temperature of the QGP created in heavy ion collisions at RHIC.
Moreover dileptons production is
mediated in the hadronic phase by light vector mesons, in particularly to $\rho$ (770), whose sufficiently small life time
(1.3 fm/c), and whose strong coupling to the $\pi-\pi$ channel makes it as prominent probe for in-medium modifications of hadron
properties.

\section{Review the results from the pre-RHIC era}
Dileptons and photons have already received attention in the
SPS program. An enhanced dilepton yield in
the low mass region (LMR: $m < 1$GeV) with respect to
the sum of known hadronic sources was first reported by CERES \cite{CER1} and
HELIOS \cite{HEL} in S+Au and S+W reactions at 200 AGeV. The discovery
triggered numerous theoretical interpretations \cite{theory} which concentrated on
the role of $\pi-\pi$ annihilation $\pi^+\pi^- \leftrightarrow \rho
\rightarrow e^+e^-$. Theoretical approaches based on in-vacuum properties of the $\rho$ systematically failed to described the data.
More successful was theoretical work with medium-modified spectral functions, including scenarios where the mass of the $\rho$ meson
\emph{drops}, according to the Brown-Rho scaling conjecture, or \emph{broadens} by strongly interacting with the hot and dense
medium. The CERES data also show that such enhancement rises significantly steeper than linearly with charged particle
density and is concentrated at very low pair-\pt, consistent with the
interpretation that the excess is due to binary annihilation processes \cite{CER2}. NA60 recently
confirmed the excess of pairs in In+In collisions at 158 AGeV
\cite{NA60_rho} and analyzed the slope parameter $T_{eff}$ extracted from the
corresponding pair-\pt spectra which rises with dilepton mass
consistent with the expectations of radial flow of the hadronic
decay source \cite{NA60_pt, Dam08}. Several recent theoretical works reasonably reproduce features like the mass distribution of the enhancement, the
centrality dependence, and the radial flow of the source. However the pair-\pt spectra exhibit a steep rise towards low pair-\pt that has not
been explained by any model so far \cite{HvH}. 

An excess of dilepton pairs has also been observed by NA38/NA50
\cite{NA50} in the intermediate
mass region (IMR: $1 < m< 3$ GeV), where the sensitivity to thermal
radiation increases. NA60 data suggest
that this enhancement can not be attributed to decays of D-mesons but may result from prompt production, as expected for
thermal radiation. The sudden drop of the slope parameter $T_{eff}$
above 1 GeV/c$^2$ in mass, advocate a partonic (i.e. non-flowing) origin
of this direct radiation \cite{NA60_therm, NA60_pt, Dam08}.
While similar kinematic window, $q_T = \sqrt{m^2+p_T^2}$ = 2 GeV, was addressed by
WA98 \cite{WA98a}
to measure direct photon spectra, the two points at low \pt extracted via HBT methods challenged the existing
predictions which could reconcile with the data only after the
inclusion of soft Bremsstrahlung \cite{WA98b} off $\pi\pi$ and $\pi K$
from the late stages of the fireball. 

\section{The new results at RHIC}
\begin{figure}
  \centering
\includegraphics[height=2.2in,width=2.5in]{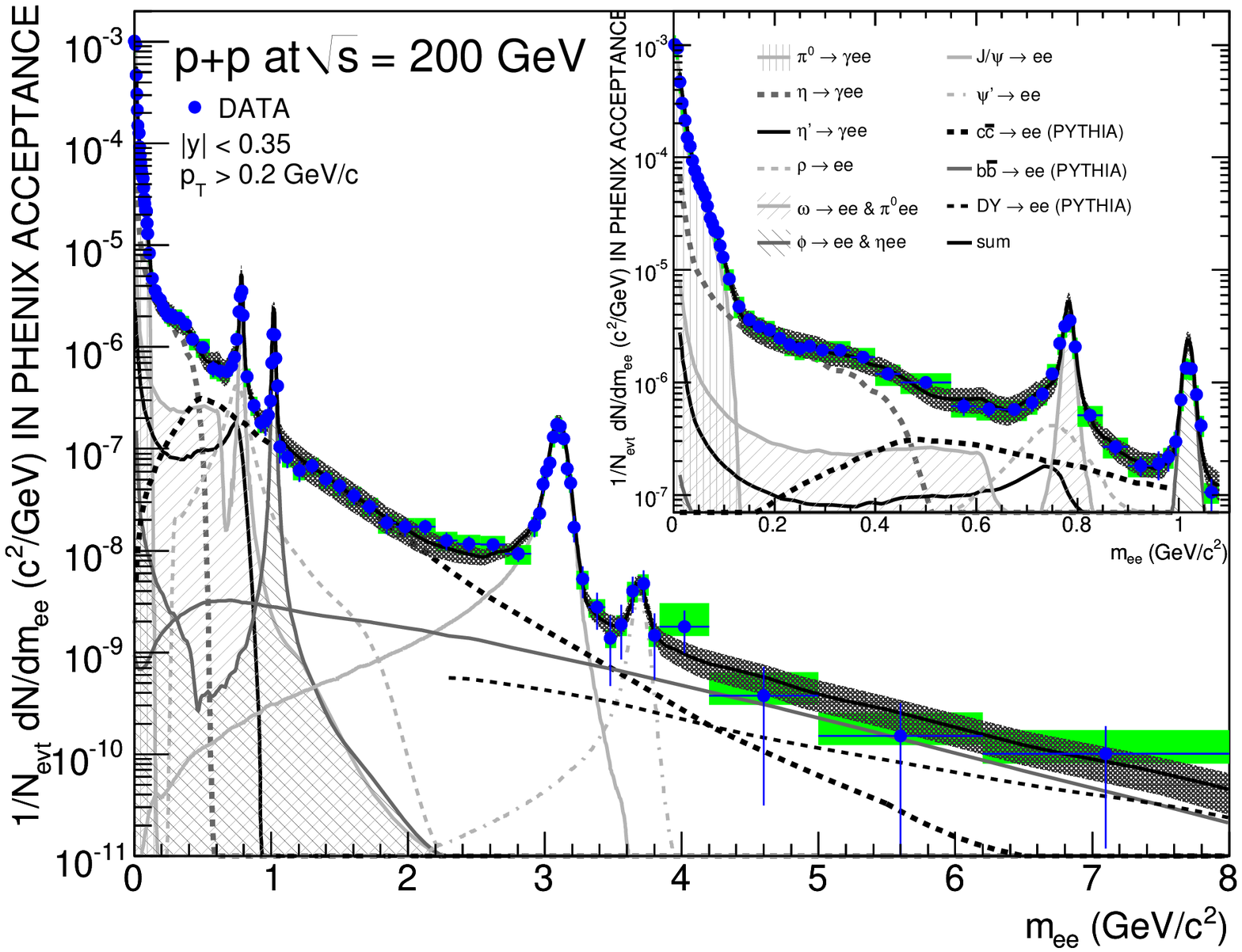}
\includegraphics[height=2.2in,width=2.5in]{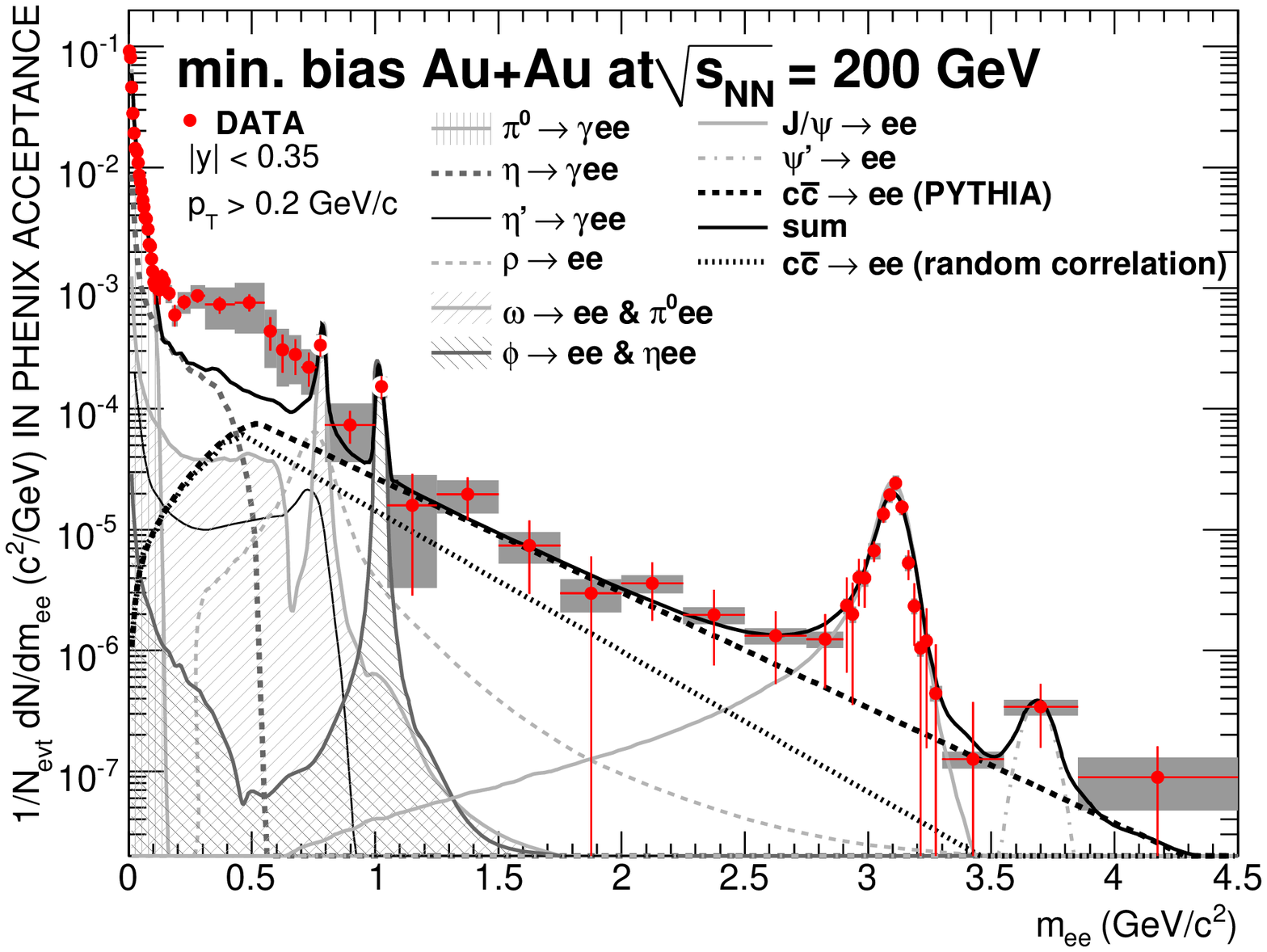}
\caption { \label{fig:cocktail}
  Electron-positron pair yield per inelastic collision as function
  of pair mass in p+p (left \cite{ppg085}) and Au+Au (right \cite{ppg075}). The data are compared
  to a cocktail of known sources. The inset in p+p shows the same data but focuses on the low mass region.}
\end{figure}

The PHENIX experiment at RHIC extends dileptons and photon measurements with the
analysis of p+p and Au+Au collisions in a new energy regime of $\sqrt{s_{NN}}$=200 GeV.
Thanks to its versatility in particle identification, high granularity
and high rate capability, PHENIX can measure electrons and photons with high precision at mid rapidity and muons at forward
rapidity.
Detailed descriptions of the electron pair analyses can be found in
\cite{ppg075,ppg085,cpod}.
The pair-signal is compared to a
cocktail of hadronic sources, semi-leptonic decays of heavy quarks and Drell-Yan.
\begin{wrapfigure}{l}{0.51\textwidth}
  \includegraphics[width=0.5\textwidth]{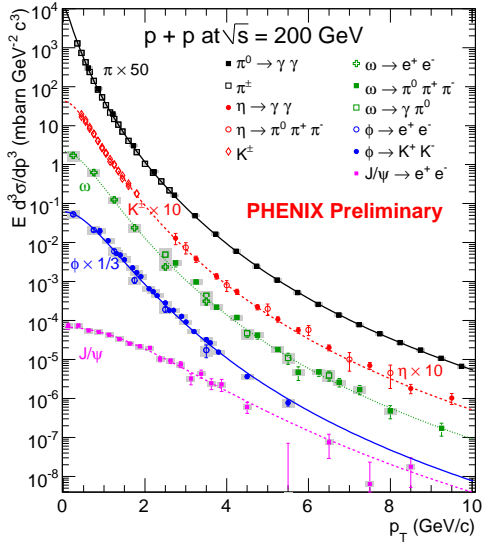}
\caption {\label{fig:mtfits}
Compilation of meson production cross sections in p+p collisions at
$\sqrt{s}$=200~GeV. Shown are data for neutral \cite{pi0} and charged pions \cite{pikp}, $\eta$
\cite{eta}, Kaons \cite{pikp}, $\omega$ \cite{omega}, $\phi$
\cite{phi} and $J/\psi$
\cite{jpsi} compared to the \mt scaling parameterization.} 
\end{wrapfigure}
In p+p collisions PHENIX has measured the spectra of pions
\cite{pi0,pikp}, kaons \cite{pikp},
$\eta$ \cite{eta}, $\omega$ \cite{omega}, $\phi$ \cite{phi}, $J/\psi$ \cite{jpsi} via several decay channels, 
see Figure~\ref{fig:mtfits}.
Pions are fitted with a modified Hagedorn function; the other data
are fitted with the same function scaled with \mt and 
a free normalization factor. Excellent agreement with the data is achieved.
Figure~\ref{fig:cocktail} shows the \ee pair yield as a function of pair
mass for p+p (left) and Au+Au (right) both compared to an absolutely normalized cocktail
tuned for the respective collision system. The
contribution from heavy quark decays has been fitted to the p+p data
using the \ee pair distribution predicted by PYTHIA \cite{pythia}. The
charm cross section was found to be
$\sigma_{c\bar{c}}=$544$\pm$39(stat)$\pm$142(syst) $\pm$200(model)
$\mu$b consistent with
QCD calculations and with more precise measurements from single
leptons (567 $\pm$57$\pm$193 $\mu$barn \cite{ppg065}).
This latter was used for the Au+Au cocktail, scaled by the average
number of binary collisions ($258 \pm 25$) \cite{pich}.
The p+p data agree very well with the sum of all known sources.

In Au+Au the data below 150 MeV/c$^2$ are well described by the cocktail of 
hadronic sources. The vector mesons $\omega$, $\phi$ and $J/\psi$ are 
reproduced within the uncertainties. However, in the mass region from
150 to 750 MeV/c$^2$, the yield is enhanced above the expectations by a factor of 3.4$\pm$0.2(stat)$\pm$1.3(syst)$\pm$0.7(model), where the first error is
the statistical error, the second the systematic uncertainty of the data,
and the last error is an estimate of the uncertainty of the expected
yield from the cocktail. Above the $\phi$ meson mass the data
seem to be well described by the continuum calculation based on
PYTHIA.
However the observation of a strong suppression and a large $v_2$ of single electrons from heavy quark
decays suggest that charm might be thermalized
\cite{ppg066}. This may correspond to a loss of the dynamical
correlation of $c$ and $\bar{c}$ quarks indicated by the second charm
curve (Fig.\ref{fig:cocktail} right), which leads to a
much softer mass spectrum and would leave room for significant other
contributions, e.g. thermal radiation.

\subsection{The centrality dependence}
To shed more light on the possible origin of the Au+Au enhancement, 
the centrality and the \pt dependence of the continuum yield are studied.
Using simulations based on a Glauber model calculation \cite{glau} the
average number of participants $N_{part}$ and
binary collisions $N_{coll}$ associated with
each centrality bin is determined.
\begin{figure}[b]
\includegraphics[height=2.2in,width=2.8in]{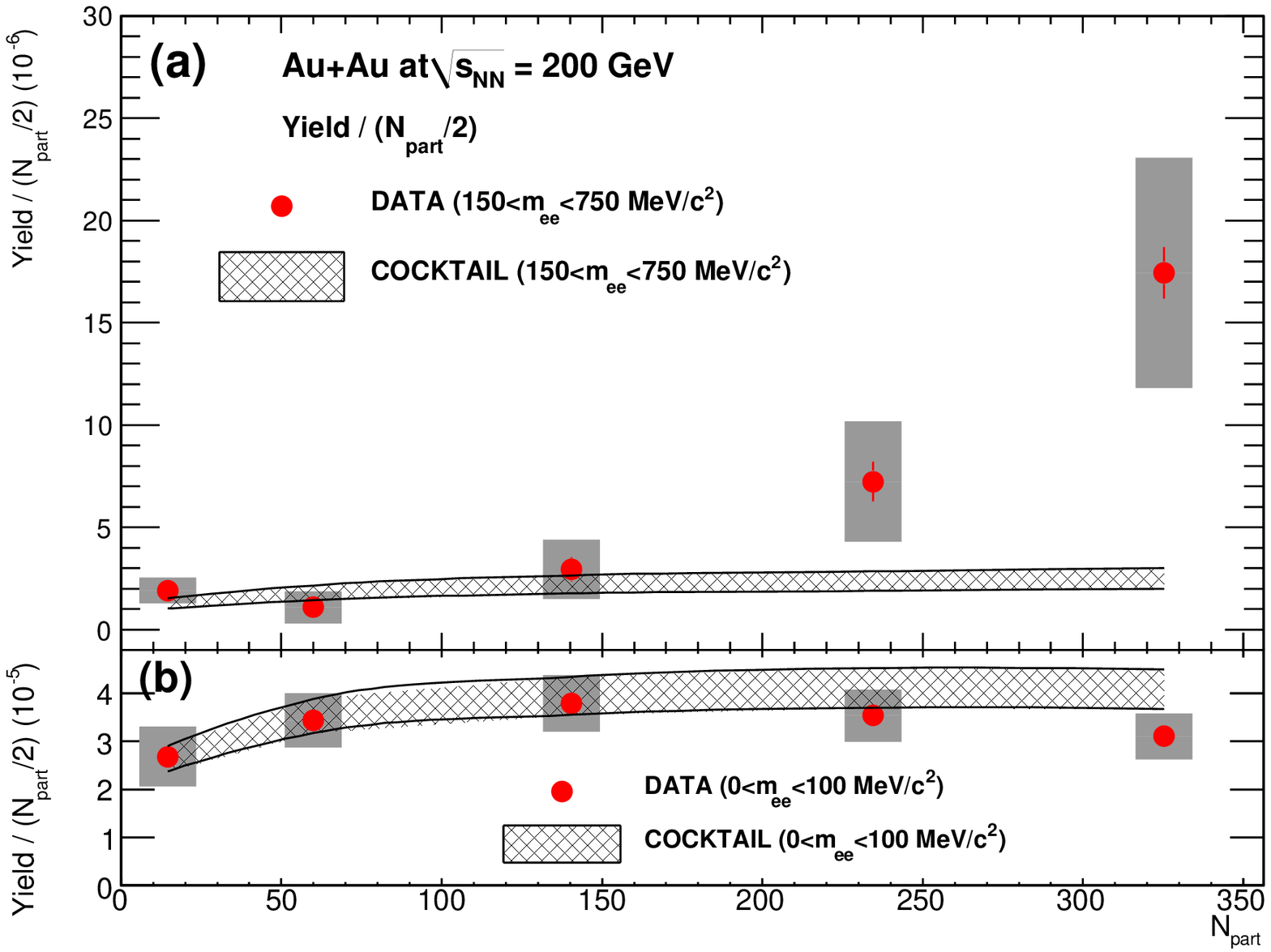}
\includegraphics[height=2.2in,width=2.5in]{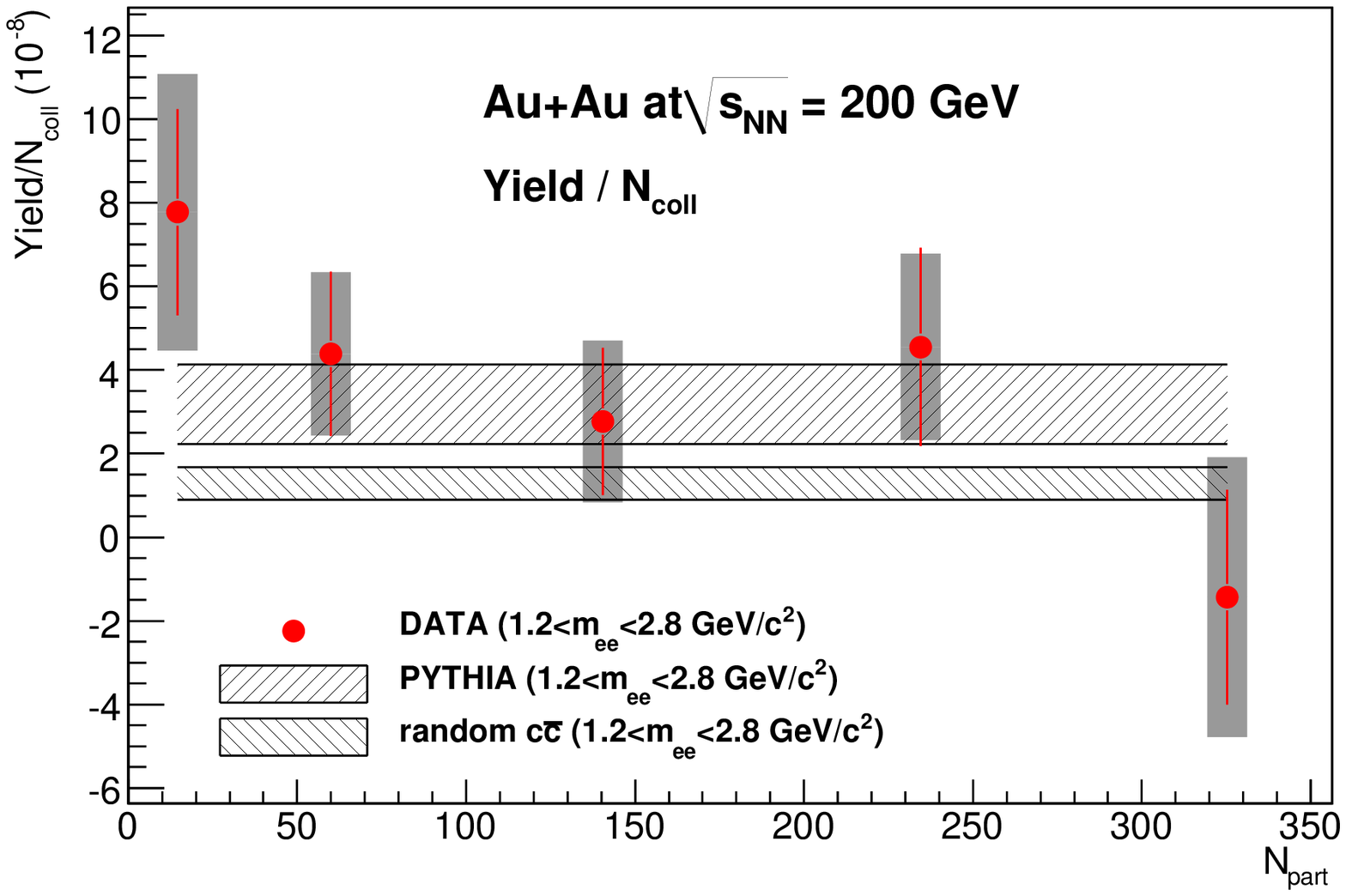}
 \caption{\label{fig:ratio}
Dielectron yield per participating nucleon pairs $N_{\rm part}/2$
(left) or per binary collisions $N_{\rm coll}$ (right) as
function of $N_{part}$ for three different mass ranges compared to the
expected yield from the hadron decay model \cite{ppg075}.}
\end{figure}
The centrality bins cover 0-10\%, 10-20\%, 20-40\%. 40-60\%, 60-92\%
fraction of the inelastic Au+Au cross section.

Fig.~\ref{fig:ratio} (left) shows the
centrality dependence of the LMR yield divided by the number of
participating nucleon pairs ($N_{\rm part}/2$).
While the yield below 100 MeV/c$^2$, which is dominated by low \pt pion
decays, agrees with the expectation from the hadron cocktail, the
yield in 150--750 MeV/c$^2$ rises significantly compared to the
expectation, qualitatively consistent with the conjecture that an in-medium enhancement of the dielectron continuum
yield arises from scattering processes like $\pi\pi$ or $q\bar{q}$ annihilation.

The yield in the mass region 1.2 to 2.8 GeV/c$^2$ normalized to the
number of binary collisions (Fig.~\ref{fig:ratio} right) shows no
significant centrality dependence and is consistent with the
expectation based on PYTHIA. However the scaling with $N_{coll}$,
expected for charmed meson decays \cite{ppg066}, may be a mere
coincidence resulting from two balancing effects: the
suppression of charm which increases with $N_{part}$ and a thermal
contribution could increase faster than linearly with $N_{part}$. 
Such a coincidence may have been observed at the SPS
\cite{NA50}, where a prompt component has now been suggested by NA60
\cite{NA60_therm}.

\subsection{The p$_\mathrm{T}$ dependence}
\begin{wrapfigure}{l}{0.75\textwidth}
  \includegraphics[width=0.74\textwidth]{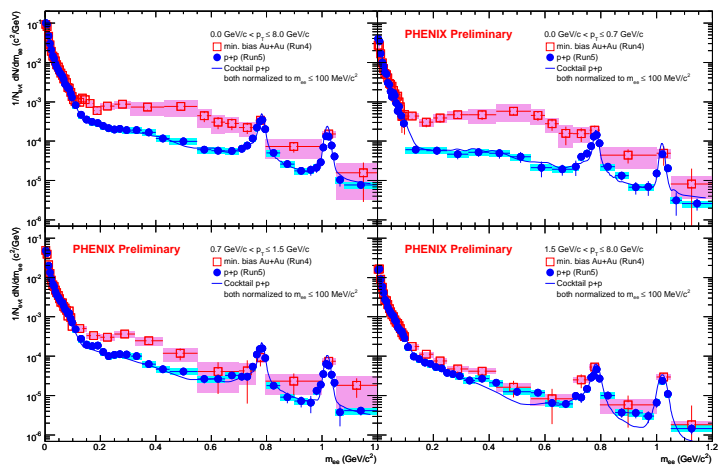}
\caption {\label{fig:mass_pp_au}
  Invariant mass spectra of \ee pairs in p+p and Au+Au
  collisions for different \pt bins.} 
\end{wrapfigure}

Figure~\ref{fig:mass_pp_au} compares e$^+$e$^-$ pair invariant
mass spectra measured in
p+p and in Au+Au collisions to the corresponding expectations from the
cocktail of hadron decays and open charm, in different ranges
of \pt.
The p+p, the Au+Au data and the cocktail are normalized relative to
each other in $m<$ 30 MeV/c$^2$. 
The p+p data is consistent with the expectations from the
cocktail over the full mass range and in every \pt bins, except for
some small deviations in the higher \pt bin (see discussion of high
\pt below).
In contrast, the low-mass enhancement in Au+Au is concentrated at
low-\pt, as shown in the Figure: the \pt bin \pt$<$0.7 GeV/c is
enhanced more than the inclusive data, the bin 0.7$<$\pt<1.5 GeV/c is
enhanced less than the inclusive data, and in the last one \pt$>$1.5 GeV/c
the Au+Au data almost agree with the p+p data.

\begin{figure}[!h]
\centering
\includegraphics[height=2.2in,width=3.2in]{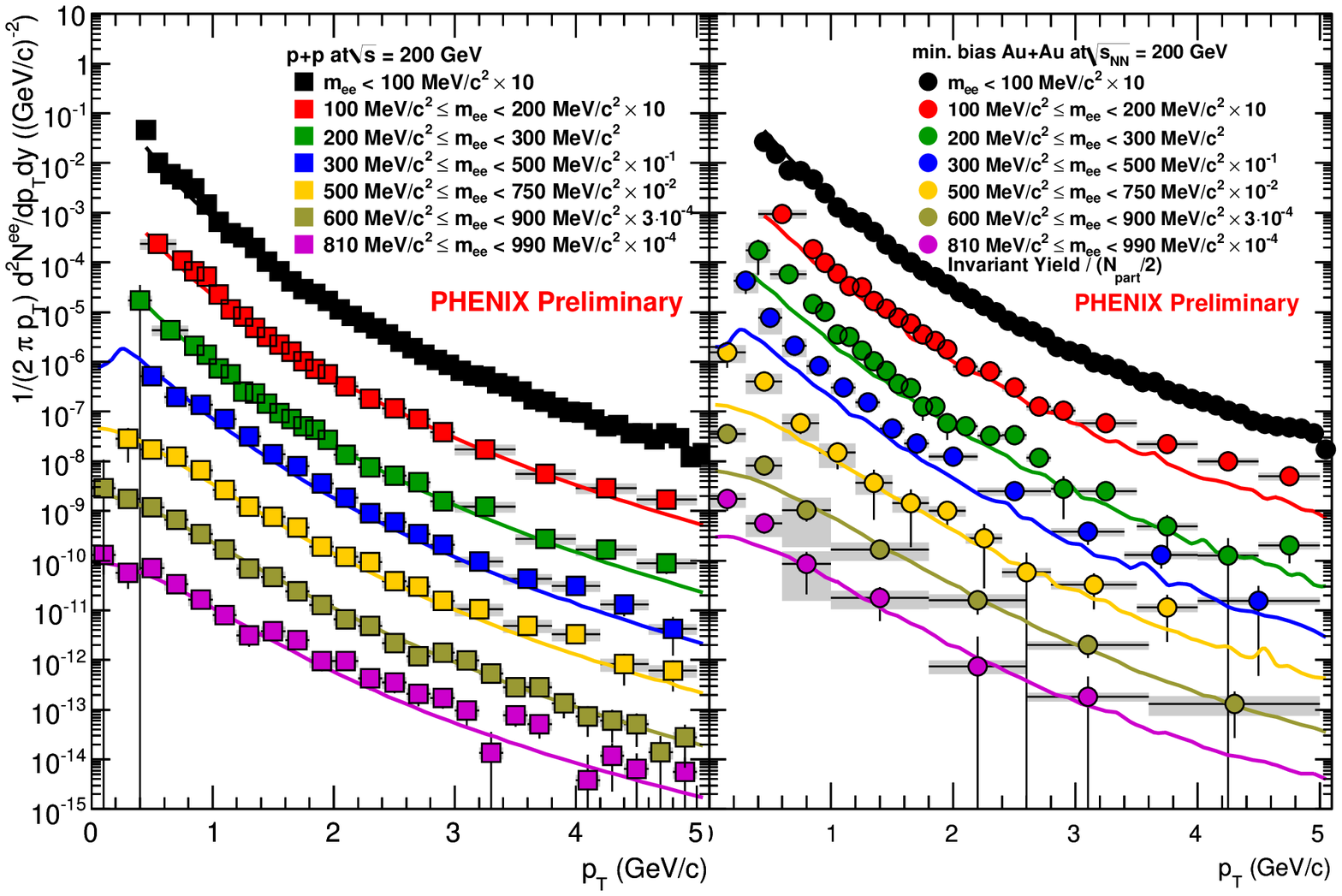}
\includegraphics[height=2.2in,width=1.8in]{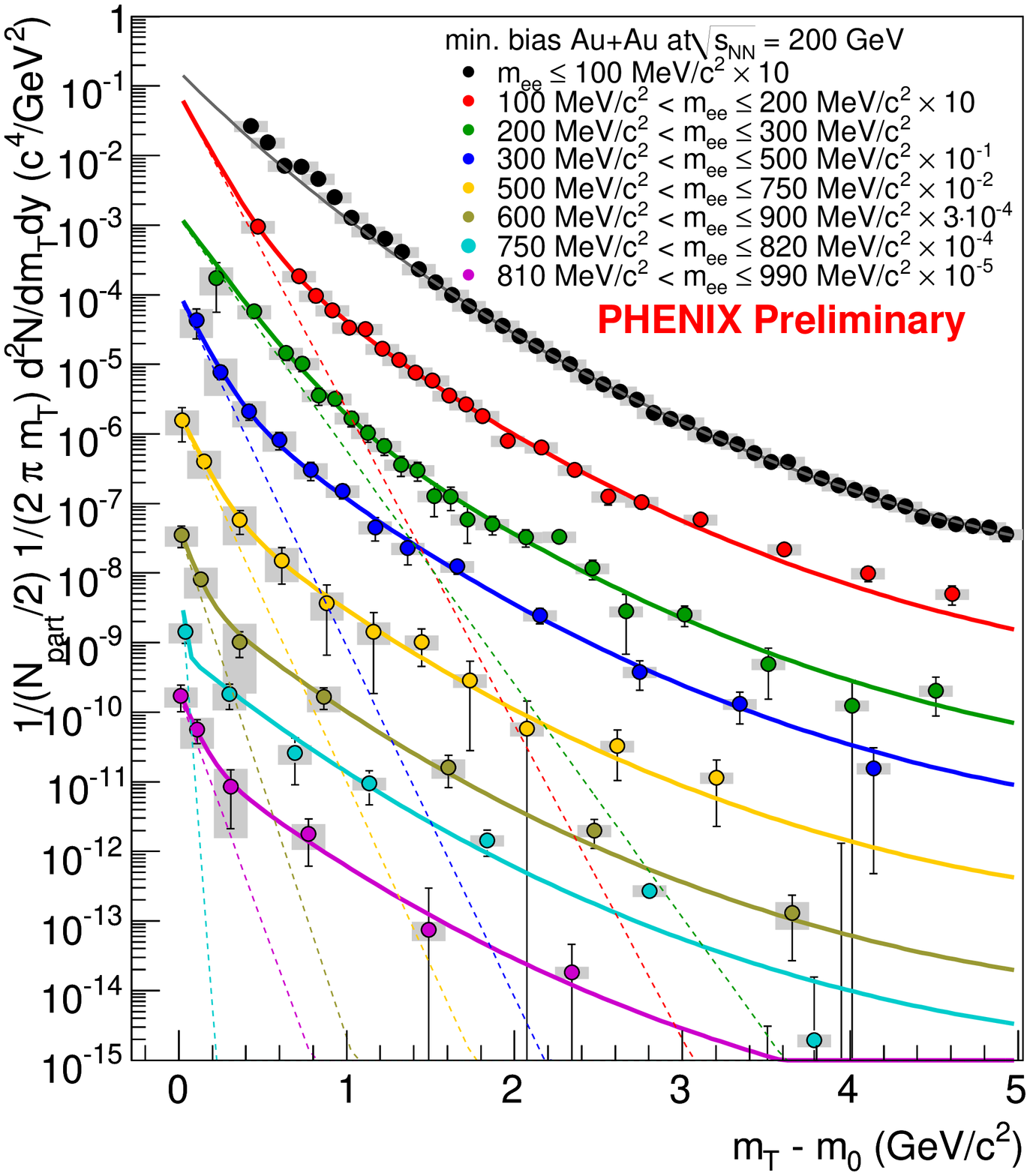}
 \caption{\label{fig:pt}
\pt spectra of \ee pairs in p+p and Au+Au collisions
for different \pt bins.}
\end{figure}

Fig. \ref{fig:pt} shows the transverse momentum spectra for different
mass windows for Au+Au and p+p data and the expected contribution
from hadronic cocktail and charmed mesons.
In the low mass bins (up to 400 MeV/c$^2$) the yield is truncated at
low pair-\pt due to the single-track \pt cut at 200 MeV/c.
In contrast to the mass spectra, the \pt spectra are corrected for the
acceptance, therefore represent the invariant yield within 
$\left|\eta\right|\le0.5$ in pseudo-rapidity and $\phi =\pi/2$ in azimuth. 
\\

In the low-\pt region (\pt$<$1GeV/c) all the p+p spectra are consistent with the expectations from the cocktail.
The Au+Au data agree with the cocktail only in the region
$m<100$MeV/c$^2$, while in all the higher mass bins the data show 
an excess that grows towards low-\pt despite the
\pt $>$ 200 MeV/c cut on the single electron track.
\begin{figure}[t]
  \centering
\includegraphics[height=2.2in,width=2.5in]{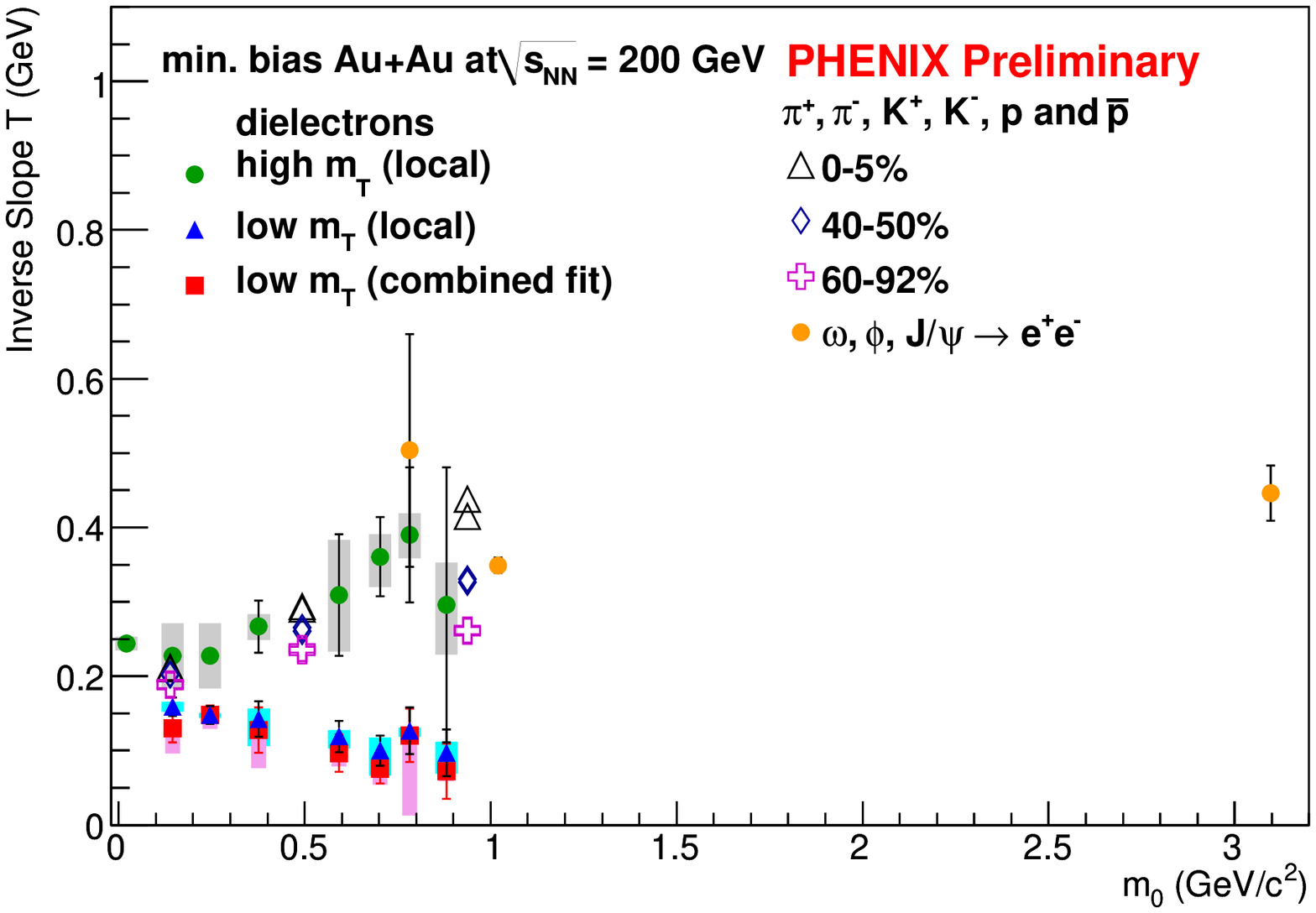}
\includegraphics[height=2.2in,width=2.5in]{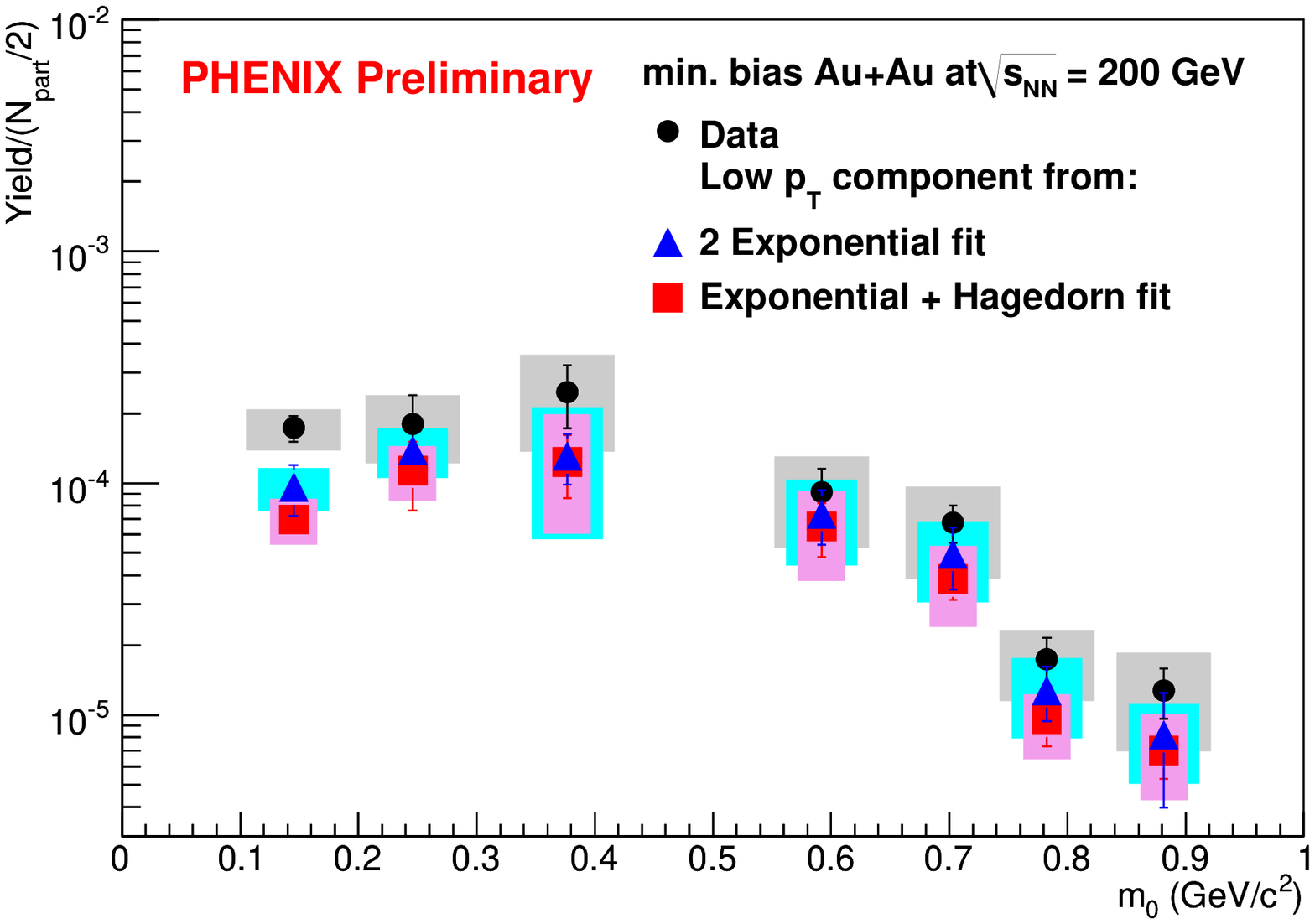}
  \caption{\label{fig:slope}
    Left: Inverse slope $T_{eff}$ of the \ee pairs and
  identified hadrons. Right: \ee pairs yield in different mass bins.}
  \end{figure}
The Au+Au spectra indicate a change of slope around 1
GeV/c. To extract information about the low-\pt component the Au+Au
spectra was fitted with an exponential in \mt in two mass regions.
The first fit is restricted to the
low-\pt region (\pt$<$1GeV/c) and the second to a region where
the data seem to follow the trend of the cocktail (1$<$\pt$<$2GeV/c). 
To account for the contribution of the cocktail below the low-\pt
component, the spectra were fitted with the sum of two exponentials;
alternatively  a modified Hagedorn function was fitted to the mass
range $m_{ee}<$0.1GeV/c$^2$ and then \mt-scaled to larger masses. 
The extracted values of $T_{eff}$ are shown in Fig. \ref{fig:slope} (left) as
a function of mass, together with the slopes of identified hadrons measured by PHENIX \cite{ppg026}.
The hadron slopes rise linearly with mass, consistent with the
expectations from radial expansion of the hadronic source. 
The slope of the \ee pair spectra in the range 1$<$\pt$<$2GeV/c, where the data are in
reasonable agreement with the cocktail, seems to follow this trend. 
In contrast, the slope of the low-\pt component remains approximately constant at $\approx$ 120
MeV and does not show the rise with mass typical for a radially expanding source. 
Fig.~\ref{fig:slope} (right) shows the \ee yield in each mass range obtained by integrating the
\pt spectra. The panel also shows the yield of the low-\pt exponential extracted from the two fits
described above (2 exponentials or exponential + \mt-scaled Hagedorn
function). In either case the yield of the low-\pt exponential
contributes at least 50\% of the yield of the spectra.
\\

\begin{figure}[!h]
  \centering
\includegraphics[height=2.2in,width=3.2in]{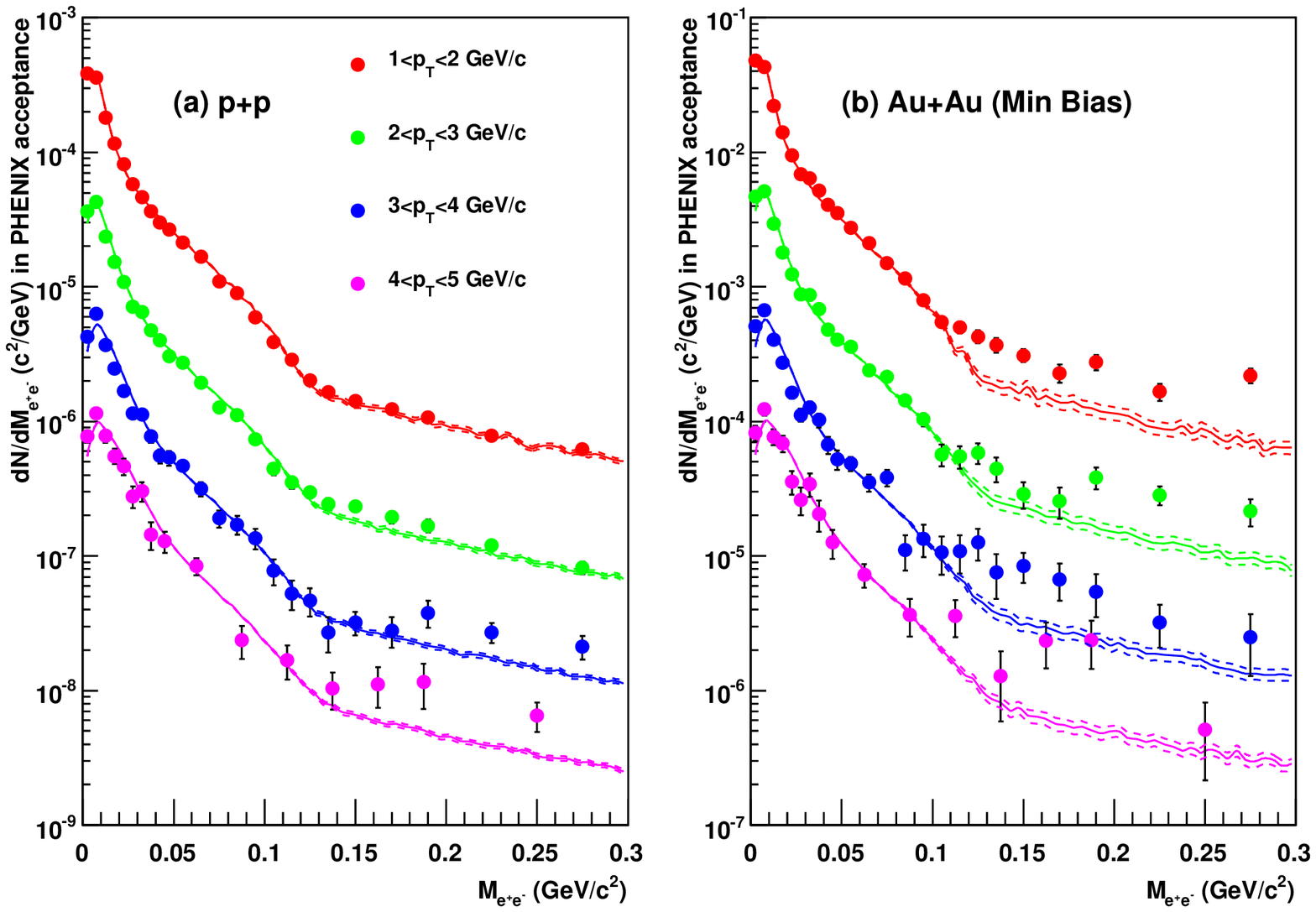}
\includegraphics[height=2.2in,width=1.8in]{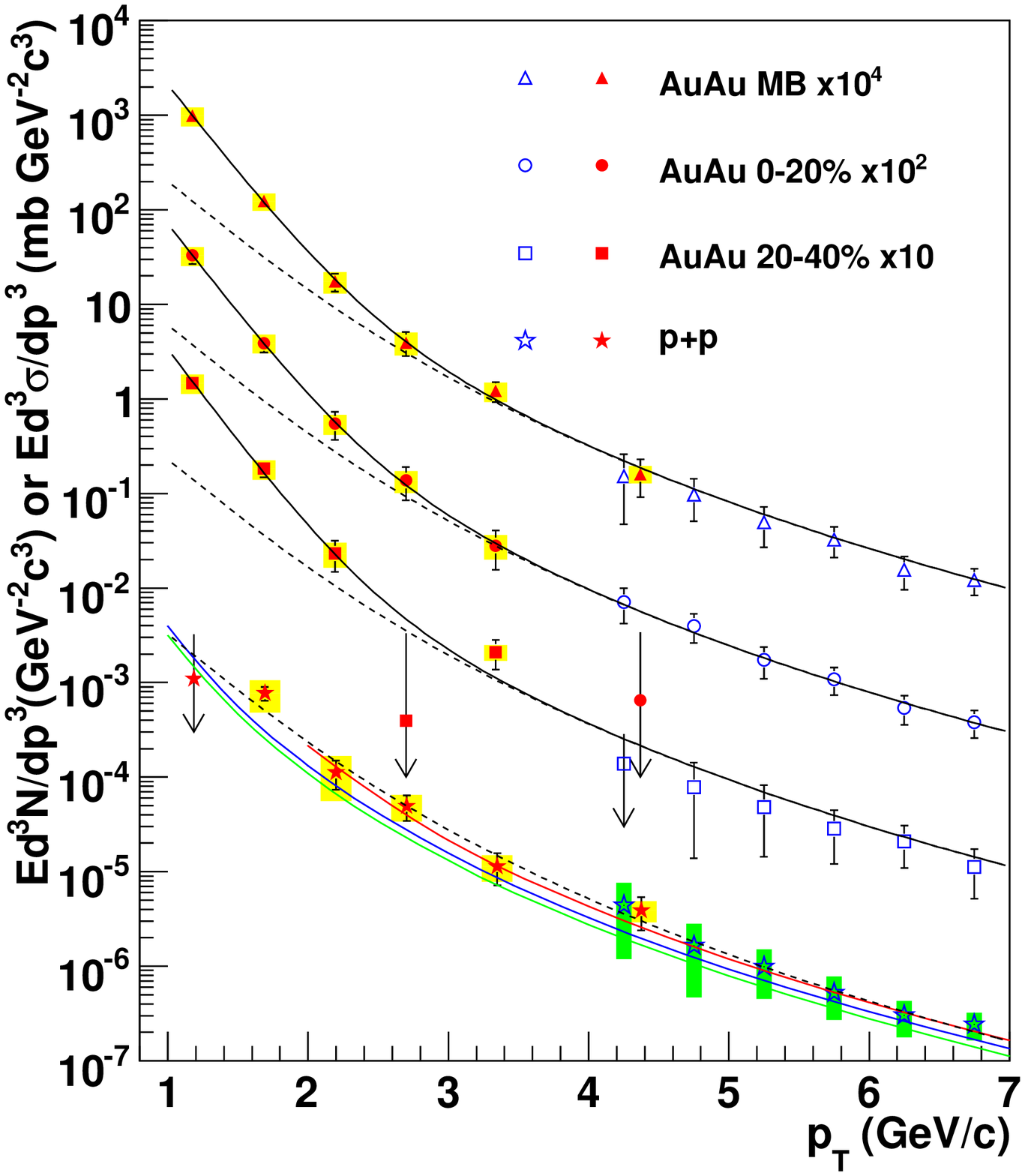}
  \caption{  \label{fig:mass_highpt}
    Left: The \ee invariant mass distributions in (a) p+p
    and (b) minimum bias Au+Au collisions compared to a cocktail of
    hadronic sources. Right: Invariant cross section (p+p) and
    invariant yield (Au+Au) of direct photons compared to a NLO pQCD
    calculation (red, green, blue curve). The black curves are
    exponential plus modified power-law fit \cite{ppg086}.}
\end{figure}

While most of the \ee yield pair enhancement in Au+Au is concentrated
\pt$<$1GeV/c, some enhancement is also observed at higher \pt.
PHENIX has also
analyzed this in the \pt region $1 < p_T < 5$ GeV/c by comparing the spectra in p+p
and Au+Au collisions to the expectations from the cocktail, as
shown in Fig.\ref{fig:mass_highpt}. 
The p+p data are consistent with
the cocktail with only a small excess in the higher \pt bins, while
Au+Au data show a much greater excess. In the mass region
$m_{ee}<300$MeV/c$^2$ little or no
contribution is expected from conventional sources 
since $\pi^{+}\pi^{-} \rightarrow e^+e^-$
can only contribute for $M_{ee} \ge 2 M_{\pi}$.
Since any source of real photons can also
emit virtual photons which convert to low mass
\ee pairs, the excess was
analyzed under the assumption that it is entirely due to internal
conversion of direct photons which result in a mass distribution
following the description by Kroll and Wada \cite{kroll-wada}.
A two-component fit which uses the relative fraction of direct photons
as only free parameter, allows to determine the fraction of direct photon component,
consistent with the expectations from NLO pQCD \cite{vogelsang} in
p+p, but larger than the $N_{coll}$-scaled calculations in Au+Au.
This is converted to the invariant cross-section/yield of direct photons,
shown in Fig.\ref{fig:mass_highpt}. 
The Au+Au data are fitted with an exponential plus a modified
power-law function fixed to the p+p data and scaled by $T_{AA}$; the
inverse slope $T = 221 \pm 23 (\rm stat) \pm 18 (\rm sys)$ is related to
the initial temperature $T_{init}$ which,
according to hydrodynamical models, is 1.5 to 3 times larger
than $T$ \cite{denterria}.

\section{Conclusions}
Measurements of Au+Au collisions at $\sqrt{s_{NN}}$=200
GeV show an enhancement of the dielectron continuum in the mass range
150--750 MeV/c$^2$ in central collisions. The enhancement is absent in
p+p and peripheral Au+Au collisions, where the observed yield agrees well
with the calculated hadron decays contributions.
The enhancement increases with centrality faster than proportional to
$N_{part}$. Moreover the enhancement is concentrated at low-\pt
with an inverse slope $T_{eff}\sim 120 MeV$ which does not exhibit a
mass dependence and thus probably does not participate in radial
collective flow.
This low-\pt enhancement is qualitatively consistent with data from
SPS experiments \cite{CER2, NA60_rho}. None of the currently available
theoretical models yet explains this low-\pt component. 
At $1<p_T<5$GeV/c the p+p data show a small excess over the hadronic
background while the Au+Au data show a much larger excess. Treating
the excess as internal conversion of direct photons, a direct photon
yield is deduced. It is consistent with NLO pQCD calculations in
p+p. In Au+Au it is consistent with hydrodynamical models with initial
temperature $T_{init}\sim 300-600$ MeV at times of 0.6-0.15 fm/c.

\def\IJMPA{{Int. J. Mod. Phys.}~{\bf A}}
\def\JPG{{J. Phys}~{\bf G}}
\def\NCA{Nuovo Cimento}
\def\NIM{Nucl. Instrum. Methods}
\def\NIMA{{Nucl. Instrum. Methods}~{\bf A}}
\def\NPA{{Nucl. Phys.}~{\bf A}}
\def\NPB{{Nucl. Phys.}~{\bf B}}
\def\PLB{Phys. Lett. B}
\def\PLC{Phys. Repts.\ }
\def\PRL{Phys. Rev. Lett.\ }
\def\PRD{Phys. Rev. D}
\def\PRC{Phys. Rev. C}
\def\ZPC{{Z. Phys.}~{\bf C}}
\def\EPJ{Eur. Phy. J. {\bf C}}
\def\etal{{\it et al.}}


\begin{thebibliography}{99}
\bibitem{pythia} We used PYTHIA 6.205 as in \cite{ppg075,ppg085}.
\end{thebibliography}
\end{document}